\documentstyle[multicol,prl,aps]{revtex}
\date{June 1st, 2000}
\input{epsf.tex}

\begin{document}

\title{Constraints on the Quasiparticle Density of States in High-$T_c$ Superconductors}

\author{T. Cren, D. Roditchev, W. Sacks, and J. Klein}

\address{Groupe de Physique des Solides, 
Universit\'es Paris 7 et Paris 6, \\  
Unit\'e Mixte de Recherche C.N.R.S. (UMR 75\ 88), 
2 Place Jussieu, 75251 Paris Cedex 5, France}

\maketitle
\begin{abstract}
In this Letter we present new tunneling data on YBa$_2$Cu$_3$O$_7$ 
thin films by low temperature scanning tunneling spectroscopy. 
Unusual peak-dip-hump features, previously reported in 
Bi$_2$Sr$_2$CaCu$_2$O$_{8+\delta}$, are also found in 
YBa$_2$Cu$_3$O$_7$. To analyse these common signatures we propose a 
new heuristic model in which, in addition to the $d$-wave symmetry, 
the gap function is energy dependent. A simple expression for the 
quasiparticle density of states is derived, giving an excellent 
agreement with the experiment. The dynamics of the quasiparticle 
states and the energy scales involved in the superconducting 
transition are discussed.
\end{abstract}

\begin {multicols}{2}

Getting to the heart of the quasiparticle density of states (DOS) in
cuprate superconductors is an active goal of current theoretical and
experimental research. Indeed, in the case of conventional materials,
the quasiparticle DOS in the superconducting state contains the key
information on the pairing. For this reason various methods, such as
angle-resolved photoemission (ARPES), tunneling (TS) and Raman
spectroscopies, as well as optical conductivity, are extensively used
to elucidate the pairing mechanism in cuprates. Among these, scanning
tunneling spectroscopy (STS), with its high spatial and energy
resolution, emerges as the technique of choice.

Using such a local probe at low temperature, reproducible vacuum
tunneling spectra have been obtained on many cuprates
\cite{Chang,Renner1995,Renner1996,Mallet,DeWilde,Cren,Miyakawa1999,WeiPRB},
in particular Bi$_2$Sr$_2$CaCu$_2$O$_{8+\delta}$ (BSCCO) and
YBa$_2$Cu$_3$O$_7$ (YBCO). However, one still cannot interpret the
experimental curves precisely. The shape of the DOS in BSCCO has a
non-trivial dependence on doping
\cite{Miyakawa1999,Renner1998bis,Miyakawa1998}, temperature
\cite{Miyakawa1999,Renner1998bis,Miyakawa1998,Campuzano,Norman1997,Norman1998,Ding,Loeser,Mandrus}
and magnetic field \cite{Maggio,Renner1998}. The state density is
characterized by its large value at the Fermi level, unusual
quasiparticle peaks and dip-hump structures at higher energies (see
Fig.1). While the $T_c$ vs doping curve is bell-shaped, the gap width
decreases linearly, as the doping increases from underdoped to
overdoped regimes, with no significant change in the DOS
shape\cite{Miyakawa1999,Renner1998bis,Miyakawa1998}. The temperature
dependence is also unconventional: The gap does not vanish at the
critical temperature and a pseudogap persists at $T>T_c$
\cite{Miyakawa1999,Renner1998bis,Norman1997,Norman1998,Ding,Loeser}.
Finally, a low temperature pseudogap remains when the superconducting
order is destroyed, within the vortex core \cite{Renner1998}, or due
to disorder \cite{Cren}. All these features are difficult to explain
with a BCS-type theory.

In this Letter we report our new low-temperature STS data on YBCO
thin films. We observe strongly pronounced peak-dip-hump structures
and a significant state density at zero bias. With previous data on
BSCCO, our observations suggest a common mechanism for these
features, linked to the superconducting state in cuprates. In order
to clarify this effect, we propose a new quasiparticle DOS based on
an energy-dependent gap function. A particular form for this function
is inferred from the data and has, in addition to $d$-wave symmetry,
a {\it single minimum} at a characteristic energy {\it near the gap
value}. Using such a simple approach, all features (wide
quasiparticle peaks, dips, humps and zero-bias) in the spectra of
both YBCO and BSCCO may be properly fitted with remarkably few
parameters. This procedure then allows one to follow the
quasiparticle states dynamics in the phase transition, including the
pseudogap. Moreover, the scaling of the dip position with the gap
value at different dopings
\cite{Miyakawa1999,Renner1998bis,Miyakawa1998} is a natural
consequence of the approach.

\begin{figure}
    \vbox to 6.5cm{
    \centering
    \epsfysize=6.25cm
    \epsfbox{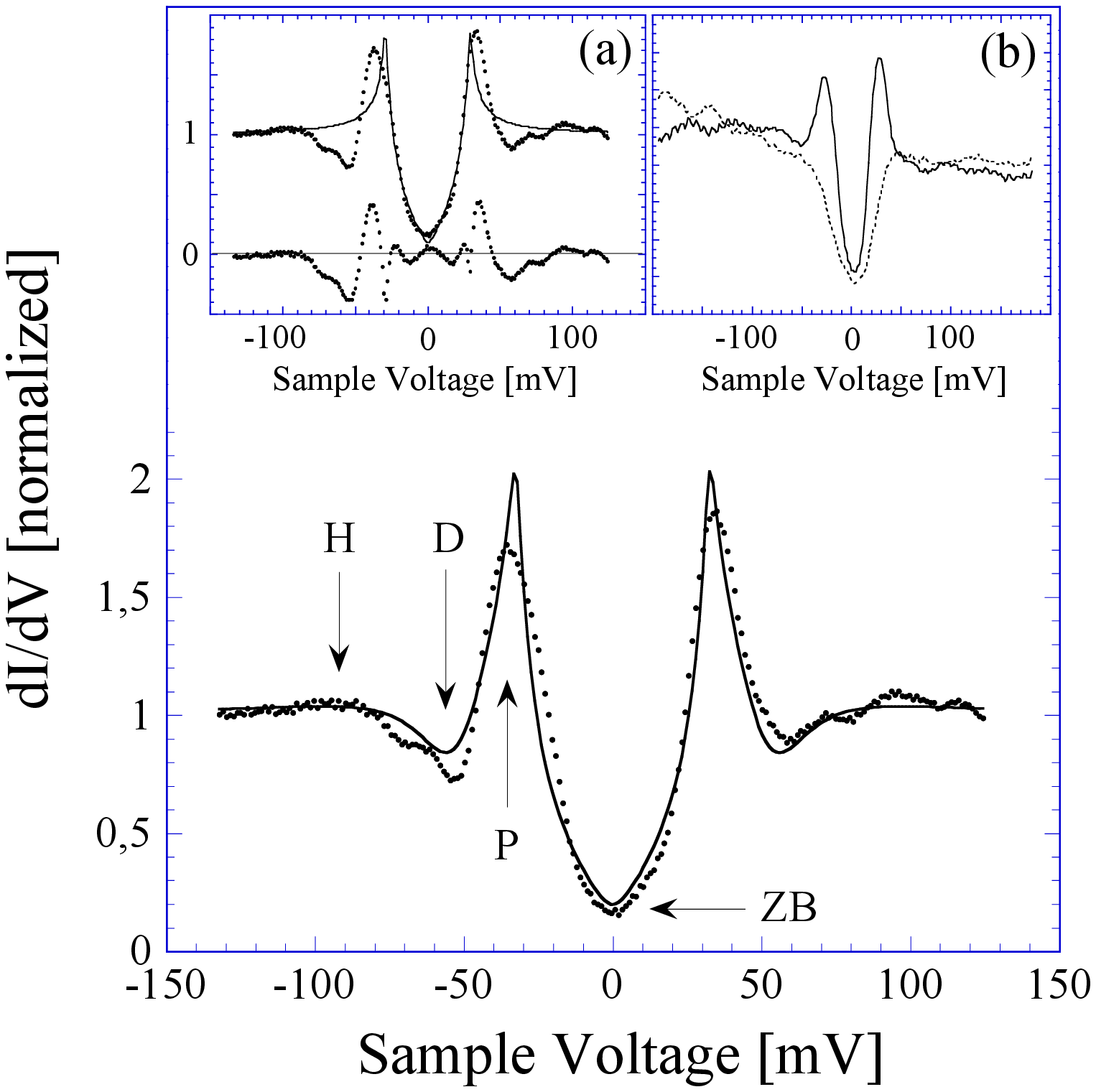}
    }
{\small
Fig.\,1. dI/dV(V) spectra on BSCCO taken at $T=4.2$ K (dots). The
positions of the quasiparticle peak, dip, hump and zero bias, are
indicated by P, D, H and ZB, respectively. Solid line: the fit by our
heuristic DOS ($\Delta_0= 46$ meV, $\delta_{0}= 17$ meV, $A_{0}=
0.45$, $\Gamma= 2.9$ meV).\\ Inset a: Experimental spectrum (dots)
and $d$-wave fit (solid line) with an anisotropic normal Fermi
surface as in \cite {Mallet}. Bottom curve: difference between
experimental and $d$-wave fit.\\ Inset b: Superconducting DOS (solid
line) and low temperature pseudogap (dotted line), measured in
disordered BSCCO at $T=4.2$ K.}\label{figure1}
\end{figure}

It is now clear that, in a first approximation, the case of BSCCO is
consistent with a $d$-wave gap function (inset (a) in Fig.1). However
the dips and humps, observed beyond the gap in the quasiparticle DOS
(Fig.1), cannot be simply explained within this approximation. These
features, existing at low temperatures, disappear at $T_c$
\cite{Miyakawa1999,Renner1998bis,Miyakawa1998}, and thus should be
related to the superconducting state. Using STS, Renner et al.
\cite{Renner1998bis} report on normal tip-superconductor (SIN)
junctions, and find the dip to scale with the gap width for different
doping levels, conserving its characteristic energy near $-2\Delta$.
This behaviour was confirmed in recent reports
\cite{Miyakawa1999,Miyakawa1998} where both
superconductor-vacuum-superconductor (SIS) and SIN junctions with
BSCCO were studied, and for a large range of carrier concentration.
In the latter work, the dips were found at both negative and positive
biases and at the energy near $\pm3\Delta$ for symmetric SIS
junctions and at $\pm2\Delta$ for SIN ones. Similar behaviour was
observed in ARPES measurements \cite{Campuzano,Norman1997}.

\vskip 1mm
\begin{figure}
    \vbox to 7.3cm{
    \centering
    \epsfysize=7.1cm
    \epsfbox{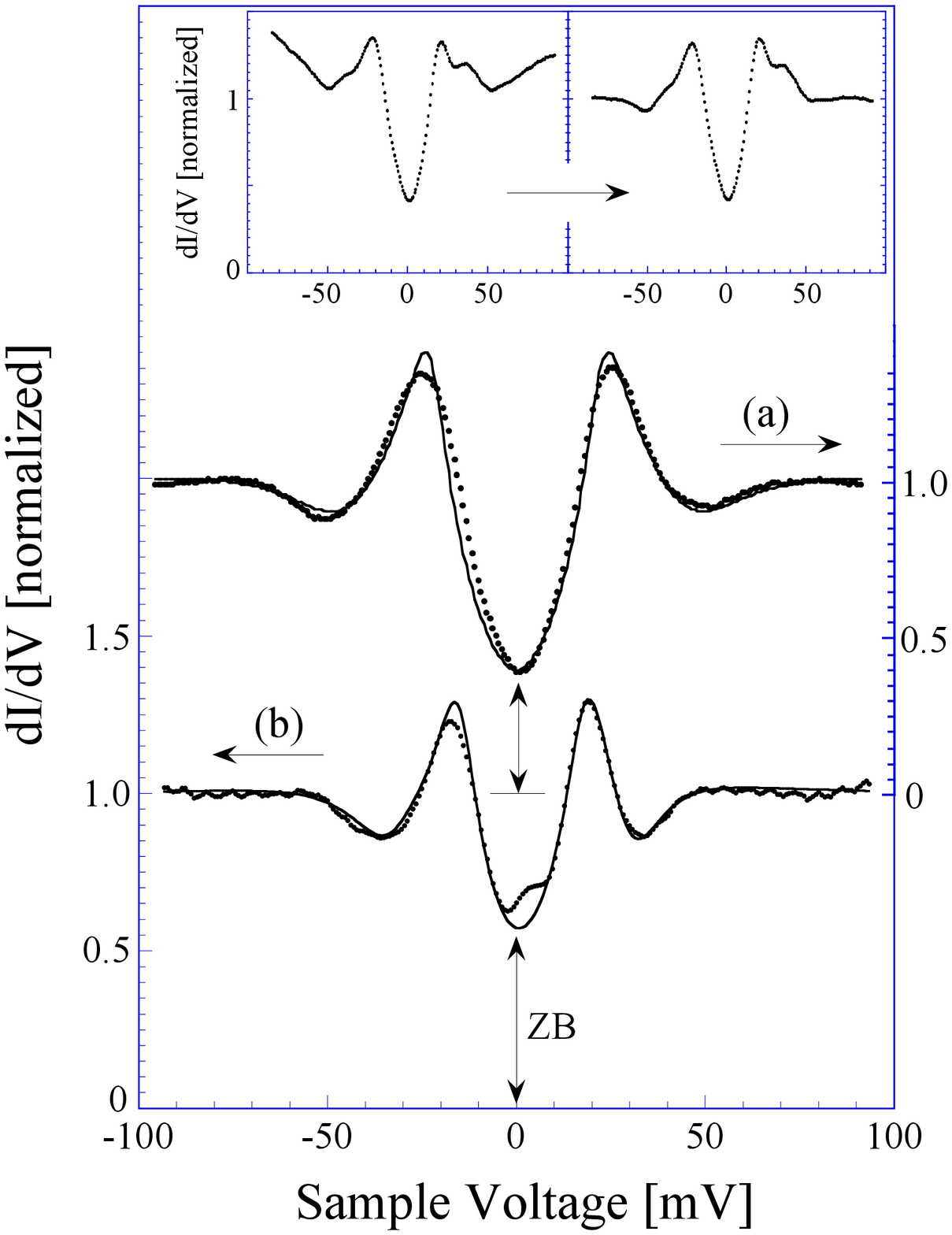}}
    
{\small
Fig.\,2. Normalized tunneling spectra of YBCO at $T=4.2$ K (dots).
Solid lines: the fit by our new DOS function ($\Delta_0= 37$ meV,
$\delta_{0}= 21$ meV, $A_{0}= 0.5$, $\Gamma= 5.5$ meV for spectrum 1;
$\Delta_0=26$ meV, $\delta_{L}= 14$ meV, $\delta_{R}= 11$ meV,
$A_{L}= 0.7$, $A_{R}= 0.45$, $\Gamma= 8$ meV for spectrum 2).
\\ Inset: Raw tunneling spectrum (left curve) observed in the same thin
film of YBCO at $T=4.2$ K. Right curve: The same spectrum, normalized
to the parabolic background.}\label{figure2}
\end{figure}

The quasiparticle DOS in YBCO was found to be different. Compared to
BSCCO, a higher density of states at the Fermi level (30-60\%
zero-bias value), a significant parabolic background and double
peaks, were observed \cite{Maggio}. However, no clear evidence for
dip signatures were reported. We indeed observe spectra of this type
in some locations of our thin films \cite{Siejka} (inset in Fig.2).
However in most of the sample surface a second type of spectrum is
found (Fig.2). The doubled peaks are absent there and a single broad
peak is followed by strongly pronounced dips at both occupied and
empty states. Another feature is an occasional anomaly at roughly +5
meV, similar to the resonant state observed in BSCCO and attributed
to a point defect \cite{Pan,Yazdani}. Comparing experimental curves, 
the shapes of our spectra in YBCO \cite{footnote} and BSCCO are quite similar. 
These data clearly show that the large peaks, the dip-hump
features and the important zero-bias background, are three signatures
of the superconducting state in cuprates, and are not just a
peculiarity of BSCCO \cite{note}. None of these features are
explained within a BCS-type weak coupling approach, even with some
particular symmetry of the gap function. We now describe a new
quasiparticle DOS, fitting in detail the tunneling spectra, and its
consequences.

Consider a dispersion law of the general form: \,$E_{\bf
k}=\sqrt{\epsilon_{\bf k}^2 + \Delta_{\bf k}^2},$ where
$\epsilon_{\bf k}$ is the normal state spectrum.  In the case of a
mean-field superconductor, the gap $\Delta_{\bf k}$ is either
constant (strict BCS) or ${\bf k}$ dependent (BCS-type). In the
latter case, of particular interest is the $d$-wave form:
$\Delta(\varphi)\simeq\Delta_{0}\,{\rm cos} (2\,\varphi)$, $\varphi$
being the angle in the {\bf a-b} momentum plane. Then, in polar
coordinates, the superconducting tunneling DOS, $N_s(E)$, is written
in terms of a spectral density $n_s(E,\varphi)$:
\begin{equation}\label{1}
  N_s(E)= \int_{0}^{2\pi} n_s(E,\varphi)\,d\varphi\ .
\end{equation}
Neglecting the angular dependence of the Fermi surface, the detailed
conservation of states $n_n(\epsilon)\,d\epsilon = n_s(E,\varphi)\,
dE$ leads to the expression:
\begin{equation}\label{2}
 n_s(E,\varphi)= n_n(E_F) \frac {E}{\sqrt{E^2-\Delta(\varphi)^2}},
\end{equation}
where $\,n_n(\epsilon)\approx n_n(E_F)=N_n(E_F)/2 \pi$ is assumed for
the normal DOS. Accounting for the anisotropy of the Fermi surface
leads to the small correction: $n_n(E_F,\varphi)$ \cite{Mallet}.
Finally, including some pair-breaking is equivalent to replacing
$E\rightarrow E-i\Gamma$ and then taking the real part. As is well
known, $n_s(E,\varphi)$ is singular at $E=\Delta(\varphi)$ and, when
integrated over $\varphi$, gives the famous $d$-wave DOS widely used
in the interpretation of the tunneling data. Having no other
singularities, expressions of this general type cannot describe the
experiment.

We go a step further by now assuming that the gap function depends on
the quasiparticle states, while the dispersion law remains unchanged.
A simple choice is of the form: $\Delta(\varphi) \rightarrow
\Delta(E_k,\varphi)$, i.e. which ignores any {\it explicit} momentum
dependence. The detailed conservation of the number of states follows
analogously, giving a new expression for the superconducting spectral
density:
\begin{equation}\label{3}
 n_s(E,\varphi)= \frac {1} {2\pi} N_n(E_F)
 \frac {E-\Delta(E,\varphi) \frac{\partial \Delta(E,\varphi) }{\partial E}}
 {\sqrt{E^2-\Delta(E,\varphi)^2}}\ .
\end{equation}
Upon integration (1), the first term of the DOS is conventional,
while the second term has a new and direct dependence on the gap
function. This result should not be confused with strong-coupling,
where the second term in (3) is absent. In fact, this formula has
been known for some time and was derived by J.R. Schrieffer
\cite{Schrieffer}. It does not apply to conventional superconductors,
in which retardation effects are crucial, and was therefore dropped.

We now focus on aspects of the spectra which determine the precise
form of $\Delta(E,\varphi)$. Recent experimental
\cite{Miyakawa1999,Campuzano,Norman1997} and theoretical
\cite{Chubukov} works have focused on the dip-hump features in BSCCO.
However, we stress that the tunneling spectra of both BSCCO and YBCO
are also characterized by unusually large quasiparticle peaks (much
larger than the energy resolution of STS at $T=4.2$ K) and an
important zero-bias background, as in Figs.1 and 2. These aspects are
even more striking if one plots the difference between the
experimental data and the $d$-wave fit (inset (a) in Fig.1). The
shape and magnitude of this difference curve lead to three
straightforward conclusions. First, the difference curve is local: It
is non-zero only within the energy range of a few $\Delta_{0}$.
Beyond these energies, there are clearly no important accidents in
the curves. Second, a simple analysis of this curve shows that the
number of states found above the $d$-wave fit is nearly equal to the
number of states found below it, as the corresponding areas are
compared. Thus, the effect is a strong but local modification of the
mean-field quasiparticle DOS. Third, the general shape of the
difference curve, in which a local maximum at one energy evolves into
a local minimum at a higher energy, indicates that they both may
arise from the derivative of a single peak at a characteristic energy
situated somewhere between.

The above arguments, together with the fact that the peak-dip-hump
appears at the critical temperature, at which the pair condensate is
formed, indicate that the mechanism responsible for these features in
the DOS should be reflected in the pair
function:\,$\Delta(E,\varphi)$. Analytically, we choose a Lorentzian
centered at $\pm |E_{0}|$ and with the characteristic width
$2\delta_{0}$:
\begin{equation}\label{4}
\Delta(E,\varphi)=\Delta_0\, {\rm cos}(2\,\varphi)\, \left[ 1-\frac
{A_0\,{\delta_0}^2} {(E\pm E_0)^2+{\delta_0}^2} \right]\ ,
\end{equation}
where $\Delta_{0}$ is the maximum value of the non-perturbed $d$-wave
gap function and $A_{0}$ is the normalized amplitude (see inset in
Fig.3). The use of (4) for $\Delta(E,\varphi)$ in (3) gives a new
quasiparticle DOS having non-trivial characteristics (Fig. 3). This
DOS has the usual '$d$-wave' singularity, but at a new energy:
$\Delta_{SC}$, defined by $E=\Delta(E,0)$. In addition, two further
extrema arise: a maximum rising near $E_{0}-\delta_{0}$ and one
minimum situated near $E_{0}+\delta_{0}$, both being due to the
derivative term.

At this stage the problem is already reduced to only five free
parameters: $\Delta_0$, $E_{0}$, $\delta_{0}$, $A_{0}$ and the
pair-breaking $\Gamma$. The fit to the experimental spectra with
these parameters gives an excellent agreement. Surprisingly, the best
fits do not situate $\Delta_0$ at the quasiparticle peak but at some
higher energy. Furthermore, the $E_{0}$ values are found to be nearly
{\it equal} to $\Delta_0$ for both BSCCO and YBCO spectra (solid
lines in Fig.1 and Fig.2a respectively). For simplicity our
calculated curves are taken to be symmetric with respect to zero
energy, whereas the experimental spectra are clearly not. However,
one can improve the quality of the fit considering $\delta_{0}$ and
$A_{0}$ to be different for the occupied and empty states in (4) (an
example is given in Fig.2b).

In order to see how far the condition $E_{0} \approx \Delta_{0}$
holds in general, we {\it a priori} set the resonance energy $E_{0}$
equal to $\Delta_0$ in all other fits. In this way, with only four
free parameters, we succeeded to fit almost all of our data.
Consulting other reports
\cite{Chang,Renner1995,Renner1996,Mallet,DeWilde,Miyakawa1999}, we
find that our quasiparticle DOS should be widely applicable. The fits
are not as good for the double peaked spectrum in YBCO, as in the
inset in Fig.2. The expression (3) allows double singularities as
shown in Fig.3d. However, they appear slightly smoother than in the
TS curves. The possible reasons may be that the Lorentzian form (4)
is too simple to reflect the orthorombicity of this cuprate, and its
smaller anisotropy. These arguments seem plausible if one compares
the tunneling data obtained in YBCO using STS \cite{Maggio} (double
peaks) and those from planar junctions \cite{Cucolo} (simple peaks).
The role of structural or surface disorder cannot be excluded either,
since in this work we observed the spectra of both types in the same
thin film of YBCO.

The physical origin of the effect is thus a new coupling at the
energy $E \simeq \Delta_{0}$ perturbing the quasiparticle spectrum in
the superconducting state. In the non-superconducting state the
interaction is either non-existant or inefficient and consequently
$A_{0}=0$, and possibly $\delta_{0}=0$. The gap in the DOS is then
$2\,\Delta \simeq 2\,\Delta_0$ (Fig.3a or Fig.3b). One may attribute
this non-superconducting state to the pseudogap seen at low
temperature \cite{Cren,Renner1998}. Indeed, from the latter papers,
where the pseudogap and superconducting gap were measured at the same
temperature (4.2 K), one finds the pseudogap to be slightly larger
than the superconducting gap (inset (b) in Fig.1). The values we find
for $\Delta_0$ are indeed larger than the peak-to-peak gap value
$\Delta_{SC}$ extracted from the spectra. They are consistent with
the low temperature pseudogap energy scale. The existence of two
'gap' scales ($\Delta_0$ and $\Delta_{SC}$ in our approach) is also
suggested in \cite{WeiPRL}, where Andreev reflection experiments
showed the gap of 28 meV, whereas the $d$-wave fit of the c-axis
tunneling data gave 19 meV for the gap in the same system.

In the superconducting state the interaction term becomes non-zero
($A_{0}>0$) and two signatures appear in the DOS: An additional peak
rises near $\Delta_0-\delta_{0}$ and a dip near
$\Delta_0+\delta_{0}$. This dynamics comes from the derivative term
in (3). Qualitatively the phenomenon corresponds to shifting states
from $\Delta_0+\delta_{0}$, to $\Delta_0-\delta_{0}$ {\it inside} the
gap $\Delta_0$ (Fig.3, from (a) to (c) and from (b) to (d)). The main
quasiparticle peak in the DOS results from the superposition of two
peaks: one at $\Delta_{SC}$ and another at $\Delta_0-\delta_{0}$, and
explains why the quasiparticle peaks appear so large when they are
not well separated (Fig.1 and Fig.2). In this Letter we do not
discuss the state dynamics in the phase transition to the pseudogap
state at $T=T_c$. This involves the explicit temperature dependence
of the parameters $\Delta_0(T)$, $\Gamma(T)$ and $\delta_0(T)$, which
are still controversial.

\vskip 2mm

\begin{figure}
    \vbox to 6.3cm{
    \centering
    \epsfysize=6.2cm
    \epsfbox{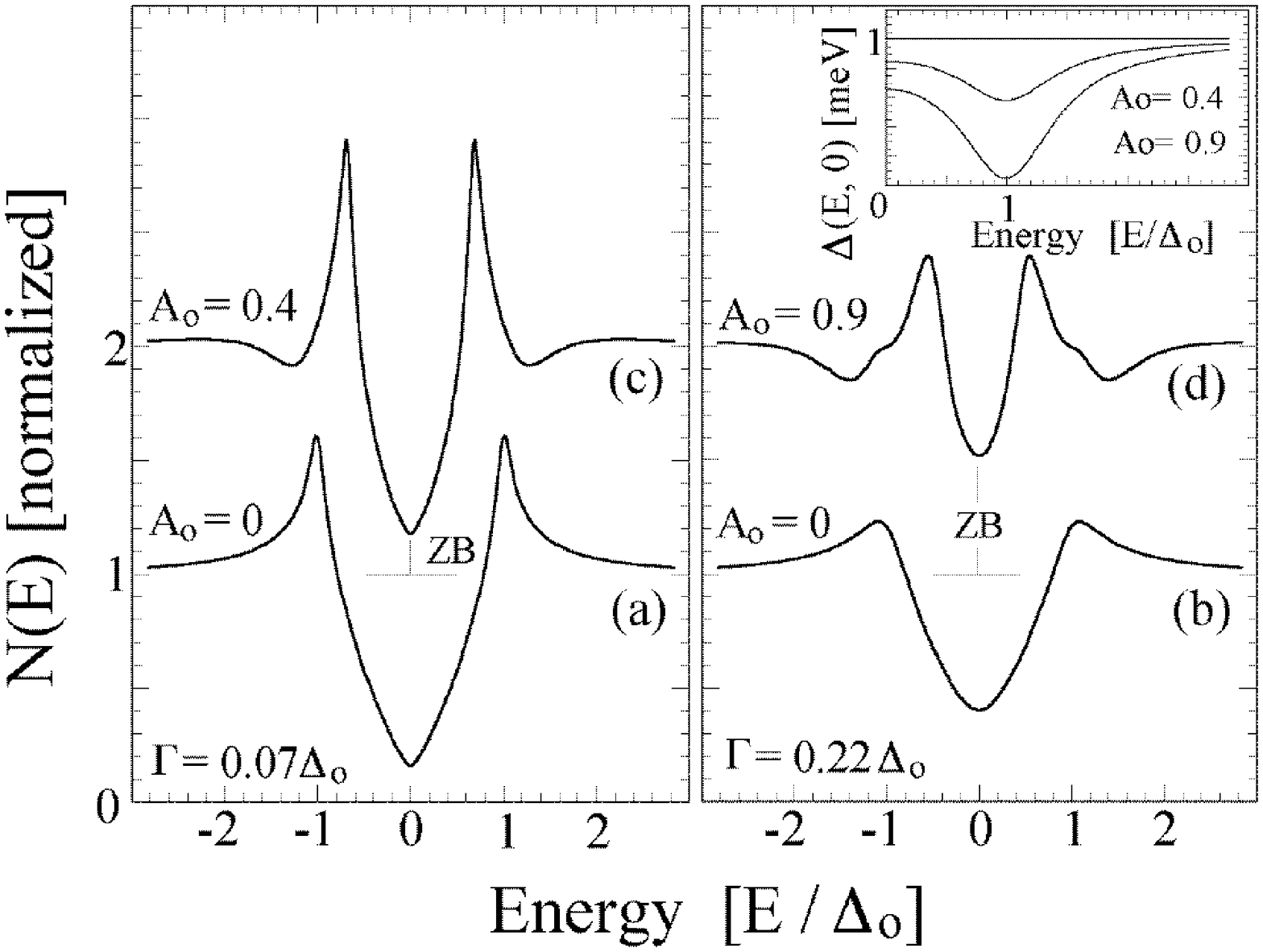}
    }
{\small
Fig.\,3. Illustration of the state dynamics from  non superconducting
DOS, (a) and (b), to superconducting DOS, (c) and (d). In (c)
$\Gamma$ and $A_{0}$ are moderate: ZB is small, the peaks are single
and well pronounced. In (d) $\Gamma$ and $A_{0}$ are large: ZB is
important, the peaks are broad and doubled, as reported in YBCO. For
all curves $\Delta_0=37$ meV and $\delta_0=17$ meV.
\\ Inset: The shapes of the energy-dependent gap function
used.}\label{figure3}
\end{figure}

As is well known, in strong-coupling approaches to conventional
superconductivity, the gap is energy dependent expressing the
electron-phonon interaction. This perturbs the quasiparticle DOS at
characteristic energies $\hbar\omega_{ph}\gg\Delta_0$ and is a second
order effect. On the contrary, our gap function is modified
significantly near $\Delta_0$, leading to new singularities in the
DOS to lowest order. In this sense they are an integral part of the
superconductivity, the relevant parameters being the amplitude $A_0$
and the width $2\delta_{0}$ of the minimum in the gap function. The
hump has no particular meaning in our model, being a consequence of
the superposition of the dip with the slow decrease of the usual DOS
tail.

In conclusion, we showed that the unusual peak-dip-hump features
exist in both YBCO and BSCCO, being common peculiarities of the
quasiparticle DOS in high-$T_{c}$ cuprates. Experiments show that
these signatures appear at $T\leq T_{c}$ and are intimately related
to the superconducting phase. We suggest that the large peaks and
dips observed in the superconducting state arise together from the
same interaction, strongly perturbing the usual mean-field DOS. This
unknown interaction is accounted for by introducing the
energy-dependent gap function, having a minimum at $E_{0}$. The
numerical analysis of the data showed that $E_{0}$ corresponds to the
maximum value of the gap $\Delta_0$ for all tunneling spectra in both
BSCCO and YBCO. The natural consequence of this is the dip-hump
scaling with the gap peak-to-peak value reported in
\cite{Miyakawa1999,Renner1998bis}. Finally, the model suggests that
the superconducting state differs from the 'normal' one by the non
vanishing of $\partial \Delta/\partial E$. These results put strong
constraints on a possible theory of the superconductivity.

We are grateful to J. Siejka and Y. Lema\^{\i}tre for providing us the
YBCO samples.

\end{multicols}


\begin{references}
\bibitem{Chang}A.\ Chang et al., Phys.\ Rev.\ B 46, 5692 (1992).
\bibitem{Renner1995}Ch.\ Renner and \O .\ Fisher, Phys.\ Rev.\ B {\bf51}, 9208 (1995).
\bibitem{Renner1996}Ch.\ Renner et al., J.\ Low\ Temp.\ Phys. {\bf105}, 1083 (1996).
\bibitem{Mallet}P.\ Mallet et al., Phys.\ Rev.\ B {\bf54}, 13324 (1996).
\bibitem{DeWilde}Y.\ DeWilde et al., Phys.\ Rev.\ Lett. {\bf80}, 153 (1998).
\bibitem{Cren}T.\ Cren et al.,\ Phys.\ Rev.\ Lett. {\bf84}, 147 (2000).
\bibitem{Miyakawa1999} N.\ Miyakawa et al., Phys.\ Rev.\ Lett.
{\bf83}, 1018 (1999) 
\bibitem{WeiPRB}J. Y. T. Wei et al., \ Phys. \ Rev. B {\bf57}, 3650
(1998)
\bibitem{Renner1998bis}Ch.\ Renner et al., Phys.\ Rev.\ Lett. {\bf80}, 149 (1998).
\bibitem{Miyakawa1998}N.\ Miyakawa et al., Phys.\ Rev.\ Lett. {\bf80}, 157
(1998). 
\bibitem{Campuzano}J.C.\ Campuzano et al., Phys.\ Rev.\ Lett. {\bf83},
3708 (1999) 
\bibitem{Norman1997}M.\ Norman et al., Phys.\ Rev.\ Lett. {\bf79}, 3506
(1997)
\bibitem{Norman1998}M.\ Norman et al.,\ Nature {\bf392}, 157 (1998).
\bibitem{Ding}H.\ Ding et al.,\ Nature {\bf382}, 51 (1996).
\bibitem{Loeser}A.\ Loeser et al.,\ Science {\bf273}, 325 (1996).
\bibitem{Mandrus}D. Mandrus et al., \ Nature {\bf351}, 460 (1991)
\bibitem{Maggio}I.\ Maggio-Aprile et al.,\ Phys.\ Rev.\ Lett. {\bf75}, 2754 (1995).
\bibitem{Renner1998}Ch.\ Renner et al., Phys.\ Rev.\ Lett. {\bf80}, 3606 (1998).
\bibitem{Siejka}The sputtered YBCO films are slightly
overdoped, $T_c=85$ K, and will be described in a future report.
\bibitem{Pan}E.\,Hudson et al., Science {\bf285}, 88 (1999).
\bibitem{Yazdani} A.Yazdani et al., Phys.\ Rev.\ Lett. {\bf83}, 176
(1999)
\bibitem{footnote}The YBCO data are normalized with respect to the parabolic part of
the spectral background, as the inset in Fig.2 shows. This procedure
does not affect the DOS at zero-bias.
\bibitem{note}In retrospect one does find dip-hump spectral features in previous works.
Unfortunately, poor tunneling conditions made their analysis
uncertain.
\bibitem{Schrieffer}J.R. Schrieffer, Rev. Mod. Phys. 200, (1964), and J.R.Schrieffer, {\it
Theory of Superconductivity}, (W.A. Benjamin, New York, 1964).
\bibitem{Chubukov}A.\ Chubukov and D.\ Morr, Phys.\ Rev.\ Lett. {\bf81}, 4716
(1998); M.\ Norman and H.\ Ding, Phys.\ Rev.\ B {\bf57}, R11089
(1998); D.\ Morr and D.\ Pines, Phys.\ Rev.\ Lett. {\bf81}, 1086
(1998).
\bibitem{Cucolo}A.\ Cucolo et al., Phys.\ Rev.\ Lett. {\bf76}, 1920
(1996)
\bibitem{WeiPRL}J.\ Wei et al., Phys.\ Rev.\ Lett. {\bf81}, 2542
(1998)

\end{references}
\end{document}